\documentclass[aps,prl,floatfix,twocolumn,showpacs,superscriptaddress,reprint]{revtex4-1}
\usepackage{amsfonts}
\usepackage{mathrsfs}
\usepackage{amsmath}
\usepackage{color}
\usepackage{graphicx}
\usepackage{bm}
\usepackage{amssymb}
\usepackage{xspace}
\usepackage{epstopdf}
\usepackage{dcolumn}
\usepackage{longtable}
\usepackage{ulem} 
\usepackage{multirow}
\usepackage[colorlinks=true, letterpaper=true, pdfstartview=FitV, linkcolor=red, citecolor=blue, urlcolor=red]{hyperref}
\usepackage[colorinlistoftodos,prependcaption,textsize=tiny]{todonotes}


\begin{document}

\title{Odd-petal states and persistent flows in spin-orbit-coupled Bose--Einstein condensates}

\author{Angela C.~White}
	\affiliation{Quantum Systems Unit, Okinawa Institute of Science and Technology Graduate University, Onna, Okinawa 904-0495, Japan}
\author{Yongping Zhang}
	\email{yongping11@t.shu.edu.cn}
	\affiliation{Department of Physics, Shanghai University,  Shanghai 200444, China}
	\affiliation{Quantum Systems Unit, Okinawa Institute of Science and Technology Graduate University, Onna, Okinawa 904-0495, Japan}
\author{Thomas Busch}
	\affiliation{Quantum Systems Unit, Okinawa Institute of Science and Technology Graduate University, Onna, Okinawa 904-0495, Japan}

\begin{abstract}

We study the  phase diagram of a Rashba spin-orbit-coupled Bose-Einstein condensate confined in a two-dimensional toroidal trap. In the immiscible regime we find an azimuthally periodic density distribution, with the periodicity highly tuneable as a function of the spin-orbit coupling strength and which favours an odd number of petals in each component. This allows for a wide range of states that can be created. We further show that in the miscible regime, both components possess states with persistent flows with a unit winding number difference between them and with the absolute values of these winding numbers depending on the spin-orbit coupling strength.
All features of the odd-petal and the persistent flow states can be  explained using a simple but effective model.
\end{abstract}





\maketitle

\section{introduction}

The recent experimental implementations of spin-orbit coupling in two-component Bose-Einstein condensates (BECs) has stimulated renewed interest in spin-orbit-coupled atomic physics \cite{Lin2011,Fu2011,Pan2012,Hamner2015}. One of the outstanding features of spin-orbit-coupled  BECs is the existence of the so-called striped ground state \cite{Wang, Ho, Sinha2011, Hu2012, Yongping2012, Martone2014,Li2016}, in which the atomic densities are periodically modulated along a spatial coordinate and the system breaks continuous translation symmetry over the scale of the BEC \cite{Li2013}.
  
Translational symmetry on the scale of BECs is usually broken by external potentials and a large number of different techniques exist to create differently shaped traps. Recently a new class of non-simply connected trapping potentials has become experimentally available, which have a toroidal or ring form. Different techniques to create these exist, which include superposing a central repulsive potential barrier onto a harmonic trap~\cite{Ryu}, hybrid magnetic-optical traps where a target ring potential is imaged onto the condensate using an intensity mask \cite{Dalibard}, or intersecting a red-detuned sheet laser and a laser using a ring-shaped Laguerre-Gaussian mode~\cite{Ramanathan, Beattie}.  Toroidal potentials allow the generation of stable and topologically nontrivial persistent atomic flow and in multi-component settings, miscibility and phase separation give rise to rich quantum dynamics \cite{Beattie,Mason, Abad, Shimodaira, White}.

The interplay between spin-orbit coupling and interactions in toroidally trapped Bose-Einstein condensates has already been shown to support unusual dynamics and tuneable features. For example, solitons propagating in Rashba-type spin-orbit coupled annular BECs can carry nonzero localized magnetization \cite{Fialko}. Previous studies have also investigated a one-dimensional combined Rashba-Dresselhaus spin-orbit coupling of a quasi-two dimensional BEC in a ring trap \cite{Karabulut}. 

In this work, we study a two-dimensional toroidally trapped BEC in the presence of Rashba-type spin-orbit coupling. We investigate the ground state phase diagram as a function of the spin-orbit coupling strength and find that in the immiscible regime rotational symmetry is spontaneously broken and the two components each form a state where the density is modulated along the azimuthal direction \cite{Xiao}. We also show that odd  numbers of petals are preferred and that the number of petals depends linearly on the spin-orbit coupling strength. In the miscible regime, both components form super-flows with different topological winding numbers.  These winding numbers depend on the spin-orbit coupling strength and in all cases a unit winding number difference exists between the two components.  All of these features originate from the spin-orbit coupling and can be physically understood via an effective one-dimensional azimuthal model.  Rashba-type spin-orbit coupling has  been realised very recently in cold atomic gases~\cite{Huang, Wu}. Our work, which describes the physics originating from applying Rashba spin-orbit coupling to BECs trapped in ring geometries, is therefore experimentally realistic and highly timely.

In the literature, two-dimensional states which are defined by a ring geometry and an azimuthally periodic density modulation are called necklace states. They
are of great interest due to their unique way of breaking a system's symmetry and were first theoretically proposed by Solja\v{c}i\'c, Sears and Segev  in a two dimensional homogenous nonlinear optical system~\cite{Soljacic1}, and were later observed experimentally ~\cite{Grow}. In homogenous systems however, necklace states are only quasi-stable and expand dynamically while keeping a necklace configuration. Stabilizing necklace states or slowing down the speed at which they expand are therefore interesting issues to address.  In recent years, necklace states have been investigated in diverse physical systems, ranging from their creation via instabilities of ring dark solitons \cite{Theocharis} or instabilities of whispering gallery modes \cite{Godey}, and as vector solutions to the cubic non-linear Schr\"odinger equation~\cite{Desyatnikov, Lashkin}, to their appearance in situations where more complex non-linearities exist. The later include examples where non-linearities of different order compete~\cite{Kartashov, Mihalache, Dong, He}, nonlocal nonlinearities exist~\cite{Rotschild, Buccoliero, Zhong, Shen} and where non-linearities are spatially modulated~\cite{Kartashov2}. So far, a stationary necklace state has only been experimentally observed in a nonlinear optical system in the presence of an optically induced periodic potential that provides a mechanism to deeply trap each petal of the necklace into a corresponding cell of the potential. The number of petals is not adjustable and is limited to eight in the experiment~\cite{Yang}.

The odd-petal states in the immiscible regime of our system stem from a complex interplay between the spin-orbit physics, which leads to the striped phase in free space \cite{Wang, Ho}, and the azimuthal symmetry of the external potential. They are therefore true two-dimensional states and we will refer to them as necklace-like.

\section{Phase diagram}

An atomic Bose-Einstein condensate with Rashba spin-orbit coupling can be described by the mean-field equations for the spinor wave function $\Phi=(\Phi_1,\Phi_2)^T$ as
\begin{equation}
i\hbar\frac{\partial \Phi}{\partial t}  = \mathrm{H} \Phi + \Gamma \Phi,
\label{GP}
\end{equation}
where the single particle Hamiltonian is given by
\begin{equation}
\mathrm{H}=\frac{p_x^2+p_y^2}{2m_{B}}  + \mathrm{V}(x,y) + \lambda (p_y\sigma_x-p_x \sigma_y),
\end{equation}
and the nonlinear part can be written as $\Gamma=\mathrm{diag}(  g_{11} |\Phi_1|^2 +g_{12}  |\Phi_2|^2,       g_{12} |\Phi_1|^2 +  g_{22}  |\Phi_2|^2  )$.
The toroidal trapping potential can be described by a shifted harmonic oscillator, $\mathrm{V}(x,y)= \frac{1}{2}m_{B}\omega_{r}^2 (r-r_{0})^2$ , where  $r^2 = x^2+y^2$. The radius of the torus is given by $r_0$, the harmonic trapping frequency by $\omega_{r}$, and $m_{B}$ is the mass of the constituent bosons.
The Rashba spin-orbit coupling strength is represented by $\lambda$, and the nonlinear mean-field coefficients $g_{11}$ and $g_{22}$ describe the s-wave scattering between atoms within each component, while $g_{12}$ represents interactions between the two components.  For simplicity, we assume that $g_{11}=g_{22}=g$ and only consider $\lambda >0$. Since the mean-field equations have a spin rotation symmetry, i.e., $U=\exp(i\pi\sigma_z/2)$, the physics in the $\lambda < 0$ regime is exactly same as that in the parameter regime of $\lambda >0$.

\begin{figure}[!]
\scalebox{0.5}{\includegraphics*{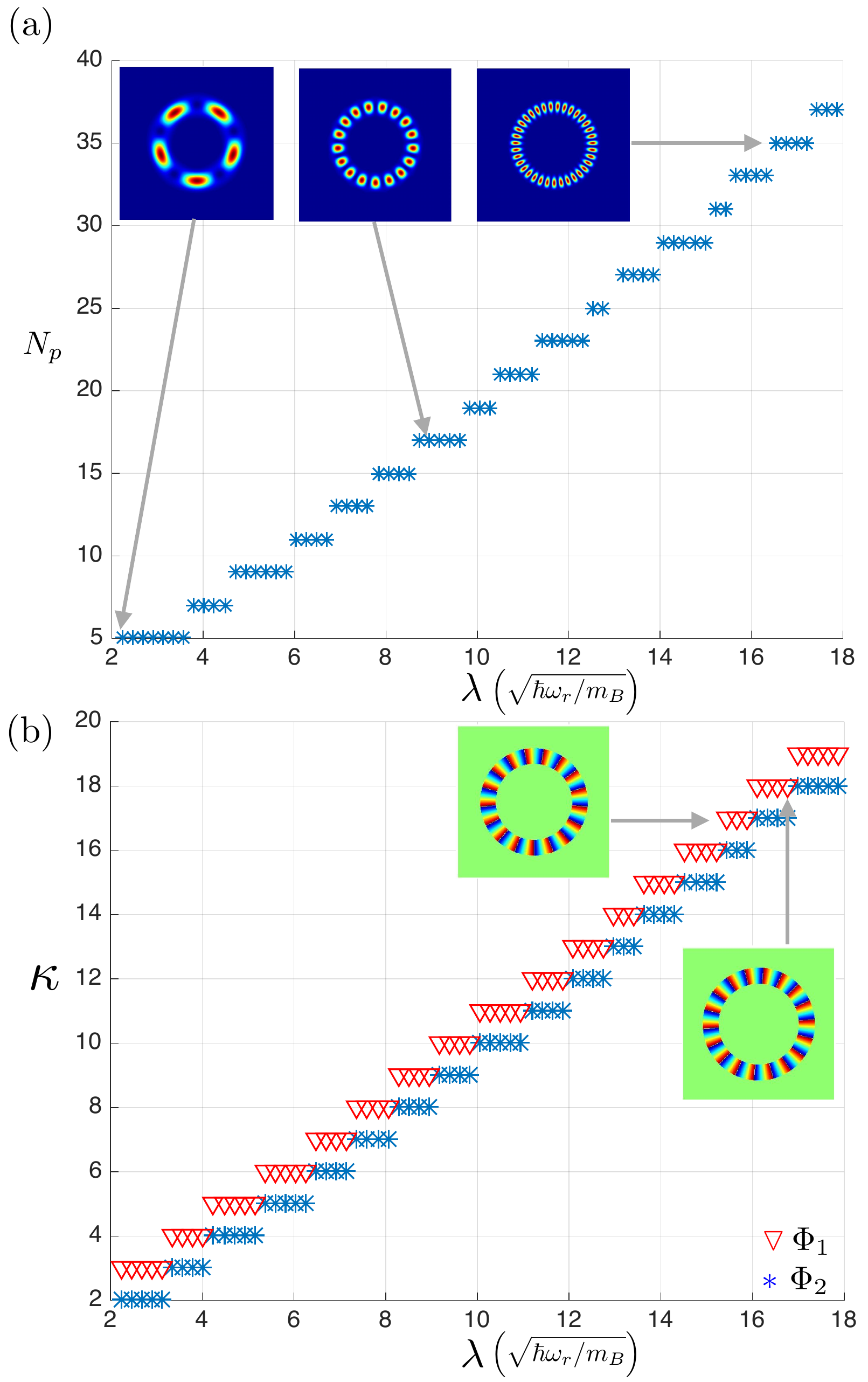}}

\caption{(Color online) (a) Number of petals as a function of the spin-orbit coupling strength in the immiscible regime $g_{12}/g=1.6$. The insets show the density profiles of a single component, confirming the typical necklace-like appearance over a large range of values of $\lambda$. (b) Winding numbers of $\Phi_{1}$ and $\Phi_{2}$ as a function of the spin-orbit coupling strength $\lambda$ in the miscible regime ($g_{12}/g=0.6$). The insets show the phase profile for $\lambda = 16.8$. Here $r_{0}= 1.12$.
}
  \label{Petals}
\end{figure}

In the following we will first numerically solve the above Hamiltonian, using standard imaginary time evolution, and then present an analytical model that confirms and explains our main numerical findings. In the numerical solutions the energy will be scaled in units of $\hbar \omega_{r}$, length in units of $\sqrt{\hbar/m_{B}\omega_{r}}$ and $\lambda$ in units of $\sqrt{\hbar \omega_{r}/m_{B}}$.
  \begin{figure}[tb]
\includegraphics[width=\columnwidth]{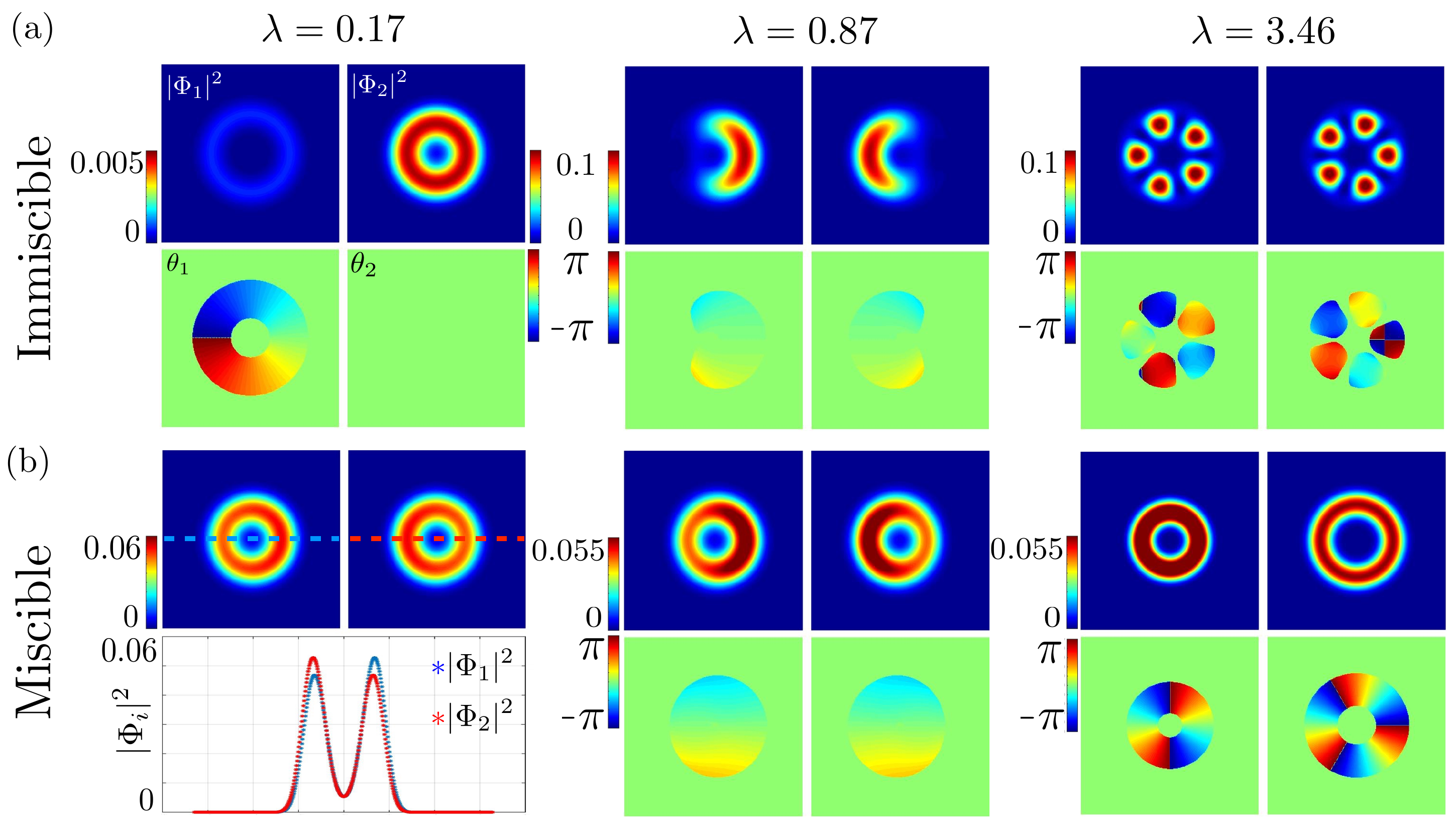}
  \caption{(Color online) Row (a) depicts examples of typical ground states of toroidally trapped condensates in the immiscible regime for  $g_{12}/g=1.2$  with increasing spin-orbit coupling strength ($\lambda = 0.17,0.87$ and $3.46$ in units of $\sqrt{\hbar \omega_{r}/m_{B}}$) and the top and bottom rows show density and phase profiles respectively.  We note that a $\lambda$ range exists between $0.87$ and $3.46$ in which the ground states consist of three petals.  Row (b) shows the miscible regime ground-states for $g_{12}/g=0.4$. The corresponding phase profiles are shown within the condensate edge (defined as $10\%$ of the maximum condensate density).  For $\lambda = 0.17$, the density profile for a slice through the centre of the condensate is shown below the full $2$D density profiles (top row of $b)$). Here $r_{0}=0.83$.
  }
 \label{Groundstates}
 \end{figure}
Typical ground states in the immiscible and miscible regimes for small and large values of the spin-orbit coupling are shown in Fig.~\ref{Groundstates}(a) and a number of interesting properties are immediately visible.
In the immiscible regime, i.e.~for  $g_{12}/g > 1$, the two components spatially separate in order to minimize the mean-field energy, which is well known from the situations without spin-orbit coupling. In the absence of spin-orbit coupling, immiscible condensates phase separate radially in wide ring traps~\cite{Mason,Abad}, or azimuthally in narrow ring traps~\cite{Shimodaira, Mason, Abad, White}. Phase separation is driven by trying to minimise the overall length of the domain boundaries, which usually leads to each component being concentrated in a single connected region.  When spin-orbit coupling is introduced, however, each of the components breaks into several parts \cite{Xiao}, 
akin to the stripe phase in simply connected condensates \cite{Li2016}.
Since this competes with the need to minimise the length of the domain boundaries, having the density stripes aligned along a single cartesian axis is not always the ground state density configuration in multiply-connected geometries. Infact, in toroidal geometries, the best compromise is to orient the stripes along the azimuthal direction, making every domain boundary as short as possible.
The stripes therefore look like petals and form a necklace-like state as shown in Fig.~\ref{Groundstates}(a) for $\lambda=3.46$.  
While it is obvious that thinner ring traps will support these azimuthal petal states, it is worth noting that this behaviour is a two-dimensional effect and does not require quasi-one-dimensionality.  As the width of toroidal traps is increased, the energy of the necklace states will get closer to the energy required for linear stripes, with the crossover determined by the competition between the strength of the spin-orbit coupling and the interaction strength between the two components, as well as the width of the ring. For fixed trapping geometry and nonlinear coefficients, the number of petals, $N_p$,  which is the same in each component, can be seen in Fig.~\ref{Petals}(a) to follow an effectively linear dependence on the spin-orbit coupling strength $\lambda$. This provides a controllable method to create necklace-like states with an arbitrary number of petals. However, we note that an odd number of petals composing the necklace-like states is dominant, which is apparent in both Figs.~\ref{Groundstates}(a) and \ref{Petals}(a). We show below that only in very narrow regimes around certain discretised values of $\lambda$  even-valued numbers of petals can be supported.

Ground states in the miscible regime, i.e.~for $g_{12}/g <1$, are shown in Fig.~\ref{Groundstates}(b). Even though we intuitively expect no phase separation between the two components, it is intriguing to note that for smaller spin-orbit coupling strengths (e.g.~$\lambda=0.17$ and $\lambda=0.87$), the ground state can be seen to be slightly phase-separated even far away from the phase transition region and become increasingly more phase-separated as $g_{12}/g$ approaches unity.  This can be understood intuitively by considering  Rashba spin-orbit coupling for a single particle and assuming the wavepacket has a finite spread in momentum space. The presence of the spin-orbit coupling then makes the momentum eigenstates of each spin component dependent on the quasimomentum, and performing a Fourier transformation from momentum space to coordinate space, this dependence shifts the wave packet center, leading to the observed displacement of the two components in coordinate space. One can easily show that this displacement is inversely proportional to the Rashba spin-orbit coupling strength, and therefore no azimuthal phase-separation exists for larger lambda.

For larger values of $\lambda$, therefore, both components exhibit homogeneous density distributions around the ring (see $\lambda=3.46$ in Fig.~\ref{Groundstates}(b)). From the phase profiles of these ground states, one can see that the winding numbers of each component can be nonzero and that there always exists a winding number difference of one unit between them.
This difference in circulation results in unequal sizes of the ring-shaped densities, and the component with larger winding sits at a larger radius in the toroidal potential due to the centrifugal barrier. Such persistent flow is very different from that in toroidally trapped rotating two component Bose-Einstein condensates without spin-orbit coupling, where the two components carry the same winding number and always exhibit homogeneous density distributions in the miscible regime.
This unit difference in winding number is maintained over a large range of values of $\lambda$ (see Fig.~\ref{Petals}(b)) and the numerically obtained ground-states indicate that the winding number of both components increases almost linearly as a function of $\lambda$.

\section{Effective model}

The exotic properties of the ground state phase diagrams identified above are unique features arising from the presence of the Rashba spin-orbit coupling. To further understand the appearance of the necklace-like states with an odd number of petals across a broad range of spin-orbit coupling strengths in the immiscible regime, and the weak phase separation and persistent flows with a unit winding number difference in the miscible regime, we will in the following use an effective one-dimensional model to provide a physical picture.

Since all of the effects highlighted above are along the azimuthal direction, we will in the following assume that the harmonic trap frequency $\omega_{r}$ is very large, so that the dynamics along the radial direction are completely confined to the lowest mode of the harmonic trap. This allows us to use a one-dimensional model, retaining only the degree of freedom along the azimuthal direction and integrating out any dynamics in the radial direction. For this we make the approximation $\Phi = R(r)\bar{\Phi}(\phi)$, where $R(r)$ is just the ground state in the harmonic potential in the radial direction, $r$.
The total energy is then given by
\begin{align}
\mathrm{E_{tot}}= \frac{1}{2\pi}\int_0^{2\pi} d \phi   \Big[      \bar{\Phi}^*  \mathrm{H_{eff}}     \bar{\Phi}
 &+\frac{\bar{g} }{2} |\bar{\Phi}_1|^4   +\frac{\bar{g} }{2} | \bar{\Phi}_2|^4  \nonumber\\
 &+  \bar{g}_{12}  |\bar{\Phi}_1|^2  |\bar{\Phi}_2|^2
\Big],
\label{TotalEnergy}
\end{align}
where $\phi$ is the azimuthal angle. The azimuthal single-particle Hamiltonian, $\mathrm{H_{eff}}$, is given by \cite{Fialko,Meijer}
\begin{equation}
\mathrm{H_{eff}} = \left( - i \frac{\partial}{\partial \phi} \right)^2  +  \bar{\lambda} \left(  \cos(\phi) \sigma_x +\sin(\phi) \sigma_y  \right) \left(   - i \frac{\partial}{\partial \phi} +\frac{\sigma_z}{2}    \right),
\end{equation}
where the energy and time units are $\hbar^2/2m_{B}r_0^2$ and $2m_{B}r_0^2/ \hbar$, respectively. Other parameters are scaled as $\bar{\lambda}=2m_{B}r_0\lambda/\hbar$ and $\bar{g}_{ij}=\sqrt{2m_{B}^3/\pi \hbar^5} r_0 g_{ij}$.  The single-particle Hamiltonian $\mathrm{H_{eff}}$ is invariant under a rotation $U=\exp(i\varphi J_z)$ due to the conservation of $J_z$, where $\varphi$ is an arbitrary angle and $J_z=-i \partial / \partial \phi  +  \sigma_z / 2 $ \cite{Ruokokoski}. This rotation invariance requires that a general ansatz to solve  $\mathrm{H_{eff}}$ should have the form
\begin{equation}
\bar{\Phi}_\mathrm{s}(m)=\exp(im\phi) \begin{pmatrix}   \bar{\Phi}_1  \\  \exp(i  \phi)  \bar{\Phi}_2   \end{pmatrix},
\end{equation}
where the integer number $m$ characterizes the phase winding induced by the spin-orbit coupling and there is always a unit winding number difference between the components.
The eigenvalues of  $\mathrm{H_{eff}}$ can be analytically found as
\begin{equation}
    E_{\pm}(m)= \frac{m^2+(1+m)^2}{2}  \pm   | m+\frac{1}{2} |  \sqrt{  \bar{\lambda}^2+1}
    \label{dispersion}
\end{equation}
and it is straightforward to see that the lower branch satisfies the degeneracy   $E_-(m) = E_-(-m-1)$.
If $\sqrt{\bar{\lambda}^2+1}/2$ is not an integer, the minima of $E_-$ appear for  $m^*$ and  $-m^*-1$, where  $m^*$ is an integer in the regime  $ \frac{ \sqrt{ \bar{\lambda}^2+1}  }{2} -1 < m^*   <  \frac{ \sqrt{  \bar{\lambda}^2+1}  }{2}$.  The lowest single particle state is therefore two-fold degenerate. However, when $ \sqrt{\bar{\lambda}^2+1}/2 $ is an integer, the minima of $E_-$ appear for  $m^*$,  $m^*-1$, $-m^*-1$ and $-m^*$, with $m^*$ being $    m^*=   \frac{ \sqrt{\bar{\lambda}^2+1}  }{2} $. The lowest single particle  state is then four-fold degenerate.

Based on these single-particles wave functions, we can now construct the ground state in the presence of the mean-field interactions. In the case of two-fold degeneracy, the trial ground state wave function is a superposition of the wave functions at the minima located at $m^*$ and $-m^*-1$ and can be written as \cite{Li, Sun}
\begin{align}
\bar{\Phi} & =    C_1 \exp(im^*\phi) \begin{pmatrix}  \cos\theta  \\ -  \exp(i  \phi)  \sin\theta  \end{pmatrix}  \notag \\
 &\qquad +C_2 \exp( - im^*\phi) \begin{pmatrix} \exp(-  i  \phi)  \sin\theta  \\  \cos\theta  \end{pmatrix},
 \label{variational function}
\end{align}
where $2\theta=\arctan(\bar{\lambda} )$ with ($ 0 < \theta < \pi/2$), which relates to the single-particle wave function at the minima. The coefficients $ C_1$ and $ C_2$ satisfy $|C_1|^2+|C_2|^2=1$ and can be determined by minimizing the total energy of the Hamiltonian in Eq.~\eqref{TotalEnergy}. Substituting the trial wave function into this expression we therefore obtain
\begin{equation}
\mathrm{E_{tot}}=\frac{2-\bar{\lambda}^2 }{2(1+ \bar{\lambda}^2)  }(g_{12}-g)|C_1|^2|C_2|^2+\mathrm{E_{con}},
\end{equation}
where $
\mathrm{E_{con}}=\bar{\lambda}^2(g_{12}-g)/4(1+\bar{\lambda}^2)+ g/2+m^{*2}+(m^*+1/2)(1+\sqrt{1+\bar{\lambda}^2}).
$
Minimizing $\mathrm{E_{tot}}$ with respect to $|C_1|^2|C_2|^2$ then leads to the following situations.

(1) For the immiscible case ($g_{12}>g$) and in the parameter regime of $\bar{\lambda}>\sqrt{2}$, one finds $|C_1|^2=|C_2|^2=1/2$. The density distributions of the ground state wave functions become
$| \bar{\Phi}_1|^2=1+\frac{1}{2}\sin(2\theta) \cos[(2m^*+1)\phi+\chi] $ and $ |\bar{\Phi}_2|^2 =1-\frac{1}{2}\sin(2\theta) \cos[(2m^*+1)\phi+\chi] $, where $\chi$ is the phase of $C_1C_2^*$, which can not be fixed by the energy minimization. It is apparent that the two components feature spatial separation along the azimuthal direction, $\phi$, and  possess equal periodicity of $2\pi/(2m^*+1)$, indicating that there are $2m^*+1$ density blocks in each component. Combining this azimuthal solution with the ground state in the radial direction leads to the appearance of spatially separated necklace-like states, with odd numbers of petals in each component, confirming the observation in the numerical simulations (see  Fig.~\ref{Groundstates}(a) and Fig.~\ref{Petals}(a)).  The state in Eq.~\eqref{variational function} with $|C_1|^2=|C_2|^2=1/2$ is analogous to the  stripe phase \cite{Wang, Ho}.

(2) In the parameter regime of $ 0<\bar{\lambda}<\sqrt{2}$ in the immiscible case, the energy minimization requires $ |C_1|^2|C_2|^2=0$, resulting in two possible states, $C_1=1$ and $C_2=0$, or $C_1=0$ and $C_2=1$, with density distributions $| \bar{\Phi}_1|^2=\cos^2(\theta)$ and $| \bar{\Phi}_2|^2=\sin^2(\theta)$ or $| \bar{\Phi}_1|^2=\sin^2(\theta)$ and $| \bar{\Phi}_2|^2=\cos^2(\theta)$, respectively. These states are spatially homogenous along $\phi$, and one of the two components dominates and is composed of the majority of atoms, as $\cos^2(\theta)=\frac{1}{2}(1+\frac{1}{\sqrt{1+\bar{\lambda}^2}})$.  This is evident in Fig.~\ref{Groundstates}(a), for a spin-orbit coupling strength of $\lambda=0.17$.
Since  $ 0<\bar{\lambda}<\sqrt{2}$, $m^*$ is fixed to $m^*=0$, which indicates that only one component has phase winding.  The existence of these spatially homogenous states seems to be counter-intuitive in the immiscible regime, where phase separation is expected. However, the large imbalance in the number of atoms in each component
and the presence of a difference in phase winding between the two components makes these states energetically preferable.

 (3) For the miscible case ($g_{12}<g$) and in the regime where $\bar{\lambda}>\sqrt{2}$, the minimization of the total energy leads to $ |C_1|^2|C_2|^2=0$. Similar to the situation considered in (2), the ground states consist of homogenous states with phase winding, i.e., one component has a winding number $m^*$ and the other has winding of $m^*+1$, or one component has winding $-m^*-1$ while the other has $-m^*$. Note that there is always a unit winding number difference between two components. Since  $ \frac{ \sqrt{\bar{\lambda}^2+1}  }{2} -1 < m^*   <  \frac{ \sqrt{\bar{\lambda}^2+1}  }{2}$, the winding numbers increase roughly linearly as a function of the spin-orbit coupling strength $\bar{\lambda}$ (see Fig.~\ref{Petals}(b)). Due to phase winding, these states carry macroscopic flows without dissipation and we call  them persistent flow states.

 (4): When $0< \bar{\lambda}<\sqrt{2}$ in the miscible regime, energy minimisation leads to $ |C_1|^2=|C_2|^2=1/2$ and one again can find states with a degree of phase separation, similar to those found in case (1). However, these are different to the phase separated states of necklace-like form in the immiscible regime.  For the miscible states one has $m^*=0$ and therefore only one petal is expected.  This behaviour coincides with the one observed in Fig.~\ref{Groundstates}(b) for $\lambda = 0.17$ and $0.87$. 
 
In the above calculations, we have used a variational method and applied a one-dimensional model to understand the features of the two-dimensional, numerical ground states. In particular, this approach explains the linear dependence of the number of petals on the spin-orbit coupling strength (as shown in Fig.\ref{Petals}(a)) and the appearance and behaviour of the winding numbers (as shown in in Fig.\ref{Petals}(b)).
Additionally, the variational ansatz confirms that the ground-states consist of an odd number of petals, $2m^*+1$. This is due to the fact that only two modes, $C_1$ and $C_2$, are assumed to play a role, since the lowest single-particle state is two-fold degenerate.
However, if $ \frac{ \sqrt{\bar{\lambda}^2+1}  }{2} $ is an integer, the lowest single particle eigenstate is four-fold degenerate. In this case the ground state trial wave function should be chosen as  a superposition of four modes, $\bar{\Phi}=C_1 \Phi_\mathrm{s}(m^*)+C_2\Phi_\mathrm{s}(-m^*-1)+C_3\Phi_\mathrm{s}(m^*-1)+C_4 \Phi_\mathrm{s}(-m^*)$, with  $    m^*=   \frac{ \sqrt{\bar{\lambda}^2+1}  }{2} $, and such a trial wave function could result in an even number of petals. However, requiring that  $ \frac{ \sqrt{\bar{\lambda}^2+1}  }{2} $ must be an integer means that only very close to certain discretised spin-orbit coupling strengths necklace-like states composed of an even number of petals can be expected, which explains why in the numerical simulations shown in Fig.~\ref{Petals}(a), only states with odd numbers of petals are visible.

\section{conclusion}

In conclusion, we have shown that Rashba spin-orbit-coupled Bose-Einstein condensates in a toroidal geometry can serve as a flexible platform to realise stable necklace-like states. We have systematically investigated the exotic ground state properties of such a system. In addition to necklace-like states, the ground states may support persistent flow. An effective one-dimensional azimuthal model was explored, and the application of a variational method was found to provide a quantitative understanding of all the salient features of the different ground state phases in the miscible and immiscible regimes.  The necklace-like states as well as states with persistent flow originate from Rashba-type spin-orbit-coupling which is two-dimensional, and they cannot exist for one-dimensional spin-orbit coupling in a ring trap \cite{Karabulut}.

Finally, it is worth pointing out that the ground states we have discussed here in the nonlinear regime do not straightforwardly follow from the ground states one would obtain in the linear limit. This can be understood by noting that in a linear, two-dimensional system, with Rashba spin-orbit coupling, an infinite degeneracy of energy minima exist, whereas in a 1D ring geometry the ground state is doubly degenerate. In either case, the superposition coefficients are unfixed and therefore a striped phase cannot be expected.

\section{Acknowledgements} 

This work was supported by the Okinawa Institute of Science and Technology Graduate University.  Angela~C. White and Thomas Busch were supported by JSPS KAKENHI-16K05461. Yongping Zhang is supported in part by the Thousand Young Talent Program of China, and Eastern Scholar Program of Shanghai.

\bibliographystyle{apsrev4-1}

\end{document}